\documentclass{llncs}

\usepackage[english]{babel}
\usepackage[utf8x]{inputenc}
\usepackage[T1]{fontenc}
\usepackage{floatrow}
\usepackage{siunitx}
\usepackage{adjustbox}
\usepackage{subcaption}
\usepackage{blindtext}
\usepackage{multirow}
\usepackage{tabularx}
\usepackage{array}
\usepackage[a4paper,top=3cm,bottom=2cm,left=3cm,right=3cm,marginparwidth=1.75cm]{geometry}

\usepackage{amsmath}
\usepackage{graphicx}
\usepackage[colorinlistoftodos]{todonotes}
\usepackage[colorlinks=true, allcolors=blue]{hyperref}
\usepackage{xcolor}
\usepackage{xspace}
\usepackage{color}
\graphicspath{ {images/} }
\title{A Comprehensive Pipeline for Hotel Recommendation System}
\author{J. Chen$^1$, Z. Gao$^{*1}$}

\institute{
$^1$Department of Computer Science, VU University Amsterdam\\
\email{vu.gaozhy@gmail.com}
}
\begin{document}
\maketitle

\begin{abstract}
    This paper addresses a comprehensive pipeline to build a hotel recommendation system with the raw data collected by Apps in users' smartphone. The pipeline mainly consists of pre-process of the raw data and training prediction models. We use two methods, Support Vector Machine (SVM) and Recurrent Neural Network (RNN). The results show that two methods achieved a reasonable accuracy with the pre-process of the raw data. Therefore, we conclude that this paper provides a comprehensive pipeline, in which a hotel recommendation system was successful build from the raw data to specific applications.
\end{abstract}

%%%%%%%%%%%%%%%%%%%%%%%%%%%%%%%%%%%%%%%%%%%
\section{Introduction}
\label{sec:int}
%%%%%%%%%%%%%%%%%%%%%%%%%%%%%%%%%%%%%%%%%%%

\par In this paper, we describe a comprehensive pipeline with two methods to predict the mood of the user on the next day based on the data we obtained from the users on the days before. Moreover, we achieve the hotel recommendation for the user based on their mood. The methods are compared with the benchmark that simply predict the mood on the next day by assuming it is the same as the previous day. First, we use the model of Support Vector Machine (SVM) to predict the mood of user. The second method we used in this assignment is the Recurrent Neural Network (RNN). Both of them achieved the prediction in a reasonable accuracy.
\par The Pre-process of the dataset is a very important step in data mining. Usually, the Pre-process is closely related to the prediction. To be clear description, this document describe the Pre-process of dataset in Section \ref{sec:pre}. The experiments are implemented in R code based on the libraries like \textit{e1071} (SVM), \textit{RNN}, etc.

%%%%%%%%%%%%%%%%%%%%%%%%%%%%%%%%%%%%%%%%%%%
\section{Pre-process the Raw Data}
\label{sec:pre}
%%%%%%%%%%%%%%%%%%%%%%%%%%%%%%%%%%%%%%%%%%%

\subsection{Data Analysis}

\begin{enumerate}
\item {\textbf {Reading and Understanding Data}}\\
\par In this section we use R code to process the dataset due to the plenty of support library in data mining. First, it is necessary to understanding what are the meaning of the variables and the value of the dataset before process the data. The dataset shows the variables and the corresponding values of the users from the smart phone. The mood of the users is related to the variables on the last days. However, the data of some variables are not related to the mood of users, or not unusable due to damage and /or insufficiency. This is the objective in this section, that we aim to pre-process the dataset from an original to the wrapped dataset that can be fed to the predictive model.\\

\item {\textbf {Pre-process the Dataset}}\\

In this section, we present the process about how we pre-process the data. First, to make the data structure clearer, we organize the dataset as the table \ref{tab:re-str} that the value is grouped by \textit{id, time, and variables}. In addition, we can analysis the mood of the user in a day like the Figure \ref{fig:histogram} for the mood dynamic of the user. In such way, we know how is the mood dynamics in a day, which is good to predict the mood of the user in the next day. Therefore we processed the dataset to the new structure as shown in Fig.\ref{fig:remove}. To build the predictive model, we need to summarise the value of the variables in days that can be right format to input the classifiers. We therefore average the value of variables in days.\\

\begin{figure}[!ht]
\centering
\includegraphics[width=0.4\textwidth]{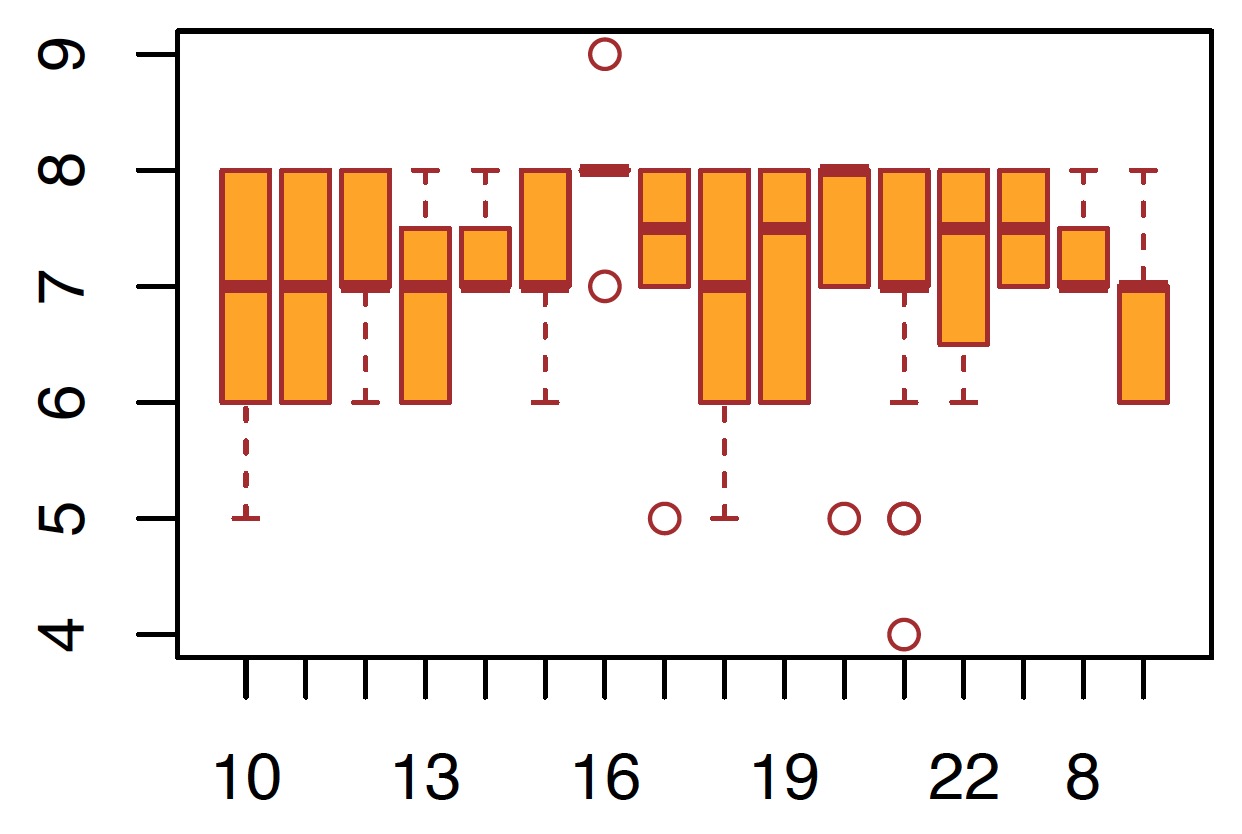}
\caption{The mood of an user in a day. The mood of the user keep in a stable average value in [7,8].}
\label{fig:histogram}
\end{figure}

\begin{figure}[!ht]
\centering
\includegraphics[width=0.5\textwidth]{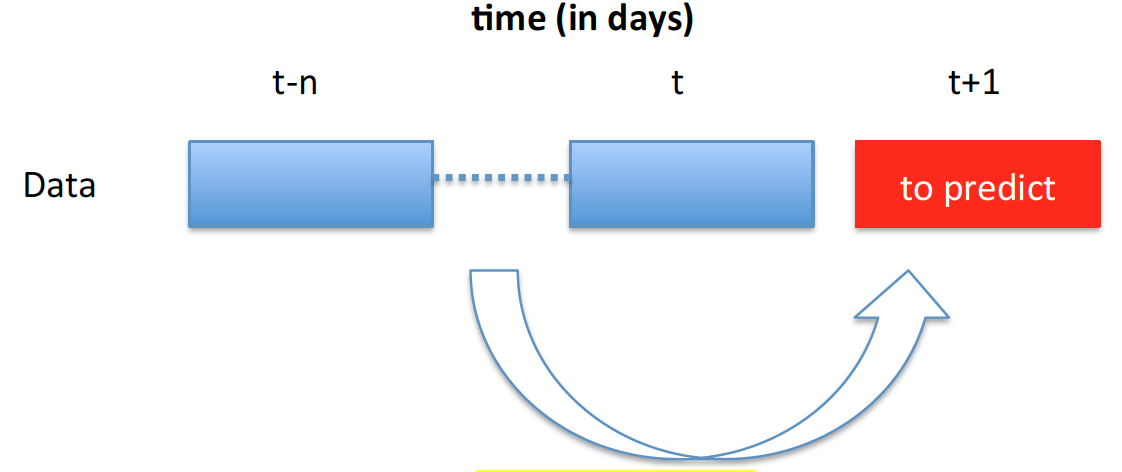}
\caption{ The predictive model procedure.}
\label{fig:predict}
\end{figure}

\begin{table}[ht]
\centering
\renewcommand{\arraystretch}{1.1}
\caption{The re-structure of dataset.}
\label{tab:re-str}
\begin{tabular}{p{2cm}p{2cm}p{3cm}p{1cm}p{1cm}c}
\hline
\multirow{2}{*}{id} & \quad \quad \multirow{2}{*}{time} & \multicolumn{4}{c}{variables}          \\ \cline{3-6} 
                    &                       & mood    & ..... & .... & ....  \\ \hline
AS14.01             & 2014-02-26            & 6.25 &       &      &  \\ \hline
AS14.01             & 2014-02-27            & 6.33 &       &      &  \\ \hline
AS14.01             & 2014-03-21            & 6.2 &       &      &  \\ \hline
AS14.01             & 2014-03-22            & 6.4 &       &      &  \\ \hline
AS14.01             & 2014-03-23            & 6.8 &       &      &  \\ \hline
\end{tabular}
\end{table}

\begin{figure}[!ht]
\centering
\includegraphics[width=0.99\textwidth]{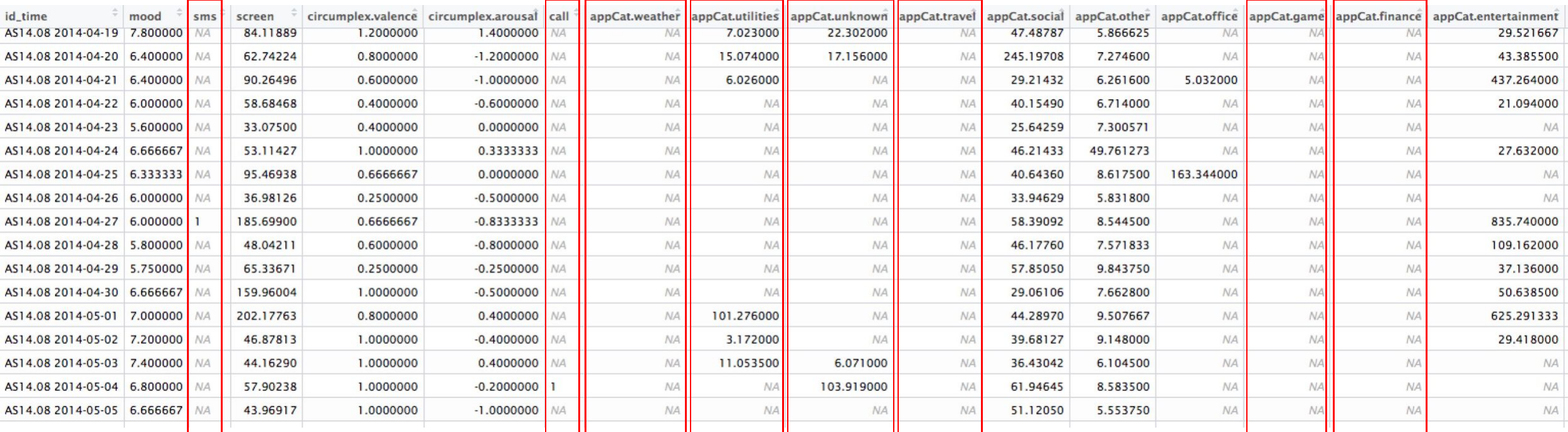}
\caption{\label{fig:remove} The snapshot of the data structure that the value is grouped by \textit{id, time, and variables}. The data in red rectangle are unusable due to too \textit{NA}.}
\end{figure}

\par However, in figure \ref{fig:remove}, we can see some variables of dataset we obtained have few data. We think they are unusable data, and remove them. Although the dataset is much more tidy, it is still not good enough to be the training data and test data. Therefore we remove the data in some days that only have a little value of variables. To here, the dataset are usable for training and testing as shown in Fig. \ref{fig:finallydata}.

\begin{figure}[!ht]
\centering
\includegraphics[width=0.99\textwidth]{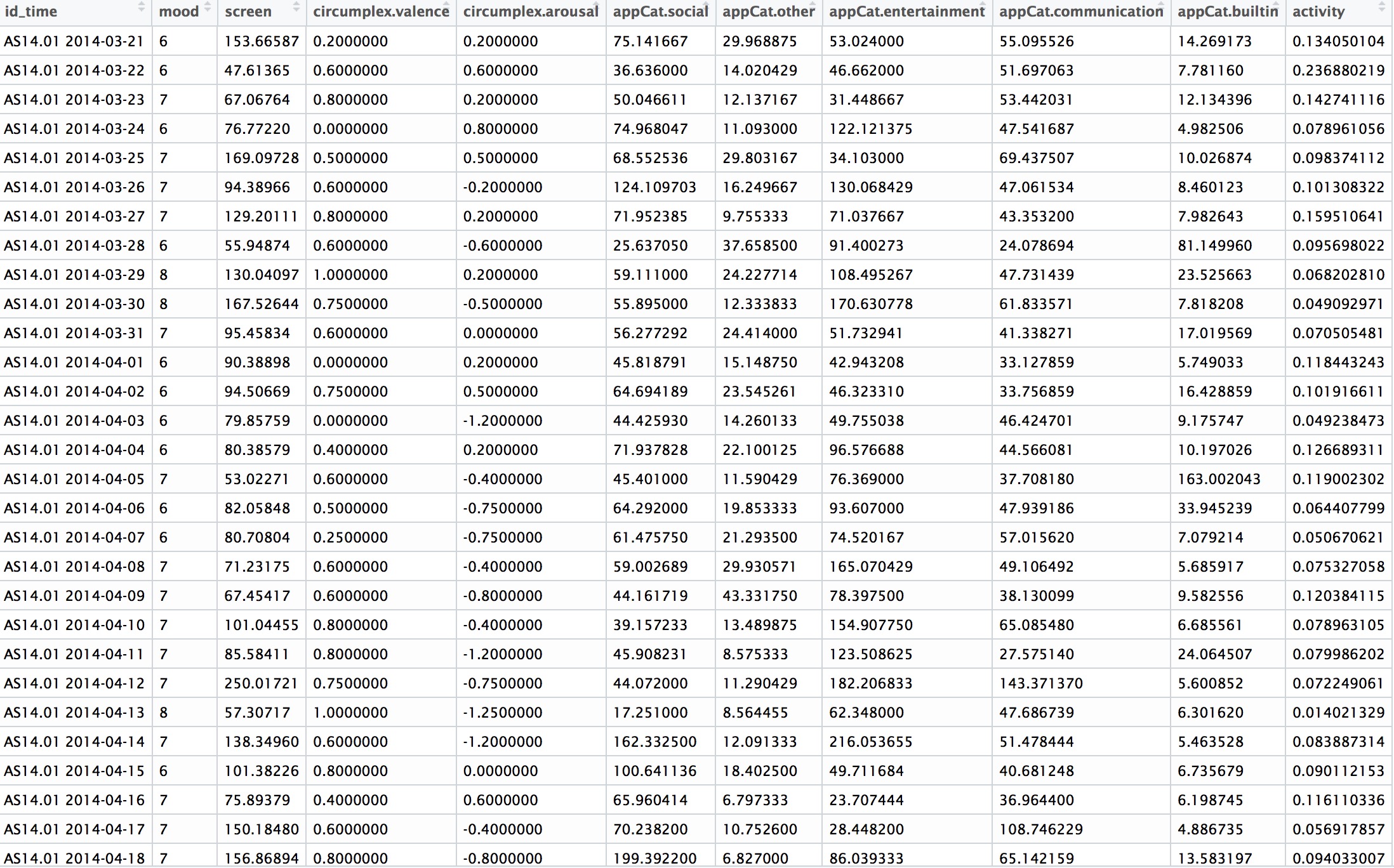}
\caption{\label{fig:finallydata} The snapshot of the usable data structure. The value is grouped by \textit{id, time, and variables}. There are no \textit{NA} and duplicated \textit{id} in different variables.}
\end{figure}

\end{enumerate}

\par To pre-process the dataset to be usable in data mining, we face many challenges and problems to the original dataset like missing value, outliers, etc \cite{che2013big}. The pre-process  is an essential and important step for data mining due to a variety of possible defect in the original dataset.

\par Here we show many examples that are part of the techniques we used in our experiments. Missing value in original dataset is a common problem that we have to solve in data mining. First, we illustrate how we process the problem of missing value in the original dataset.

\subsection{The set-up of feature}

In Fig. \ref{fig:finallydata}, we choose some variables as the usable feature with enough samples. The data in Fig. \ref{fig:finallydata} have the full values in the variables in different dates and ids. They are tidy data that can be fed to the model for training and testing from the data formate and information. 
\par
\begin{figure}[!ht]
\centering
\includegraphics[width=0.6\textwidth]{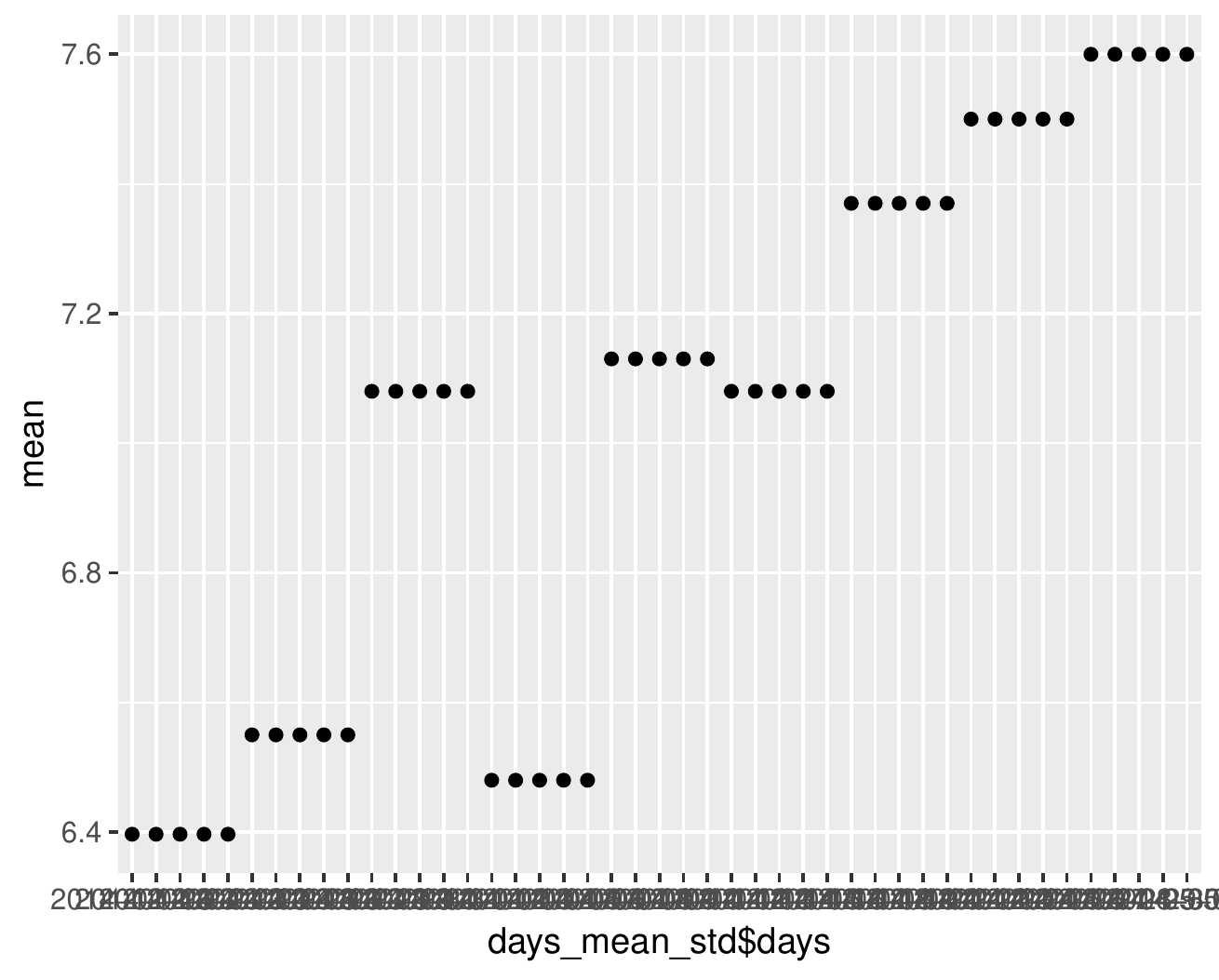}
\caption{The example of temporal abstraction. We take the average of mood value in five days.}
\label{fig:temporal}
\end{figure}

 In other hand, we divide the dataset into two parts with 10\% and 90\% rate as the training sample and testing sample separately. To build the predictive model as shown in Fig. \ref{fig:predict}, we aggregate the history to create attributes that can be used in the machine learning approach like the SVM \cite{lan2018ICARCV} and RNN we use in this document. We use the average mood during the last five days as a predictor. This is clearly present in Fig. \ref{fig:temporal} to create the new feature that can be used in the classifiers. 

\subsection{Rationale}

For the rationale of choice of the final attributes, in this assignment, we mainly consider the quality and quantity of dataset. We have to filter damaged data that probable to train an incorrect predicted model or decrease the accuracy of the prediction. Therefore, we remove the data in the day that many variables have missing value and the variable with outliers.    

%%%%%%%%%%%%%%%%%%%%%%%%%%%%%%%%%%%%%%%%%%%
\section{Learn prediction models}
\label{sec:learn}
%%%%%%%%%%%%%%%%%%%%%%%%%%%%%%%%%%%%%%%%%%%

\par In this section, we described two predictive models and the benchmark. But we are not focus on the details of the models because we use the standard R code library for SVM and RNN.

\subsection{Model Variant 1}

First, we adapted the Support Vector Machine (SVM) as the predictive model. In R programming, a variety of libraries can be used to implement SVM, we used the library \textit{e1071} due to its feature of easy-to-use. The main parameters setting of SVM is shown in Table \ref{tab:parameter_svm}.
\begin{table}[ht]
\centering
\begin{tabular}{p{1.8cm} p{0.8cm} p{2.3cm} p{1cm} p{1cm} p{1cm} p{0.8cm} p{0.7cm} p{1.9cm} p{1cm}} \hline
parameters     & scale & type & kernel & degree  & gamma  & coef0  & cost & class.weights & epsilon
\\ \hline
setting       & 1 & C-classification & linear & 3  & 1  & 0  & 1 & 1 & 0.1\\ \hline
\caption{The main parameters setting of SVM.}
\label{tab:parameter_svm}
\end{tabular}
\end{table}

Therefore, we just need to train and test the sample after the pre-process section. We used the variables in section \ref{sec:pre} as the input of SVM model and the value of mood is the output of SVM model. The 90\% sample was used to train the SVM model. The rest 10\% sample was used to test the accuracy of trained SVM model. 

\par We output the accuracy of trained SVM model by predicting the training sample. And then, The accuracy was verified again by predicting the testing sample. The table \ref{tab:train} shows the results of predictive mood of the user on the next day that testing on the training sample. In table \ref{tab:train}, the 568 samples are used to train the SVM predictive model. The value of mood from 5 to 8 are predicted from 3 to 9. The 467 samples are correct predicted. The trained SVM model have $accuracy = 467/568 = 82.2\%$.

\begin{table}[ht]
\centering
\caption{The results of training. The 568 samples are used to train the SVM predictive model. The value of mood from 5 to 8 are predicted from 3 to 9. The 467 samples are correct predicted. The statistical results in Fig. \ref{fig:predict_1}.}
\label{tab:train}
\begin{tabular}{ p{2cm} p{1cm} p{1cm} p{1cm} p{1cm} p{1cm} p{1cm} p{1cm}}
\hline
results\_train & 3 & 4 & 5 & 6  & 7  & 8  & 9 \\ \hline
5              & 0 & 1 & 4 & 1  & 0  & 0  & 0 \\ 
6              & 1 & 2 & 9 & 71 & 11  & 4  & 0 \\ 
7              & 0 & 0 & 0 & 23 & 313 & 31 & 2 \\ 
8              & 0 & 0 & 0 & 0  & 13  & 79 & 0 \\ \hline
\end{tabular}
\end{table}

\begin{figure}[!ht]
\centering
\includegraphics[width=0.5\textwidth]{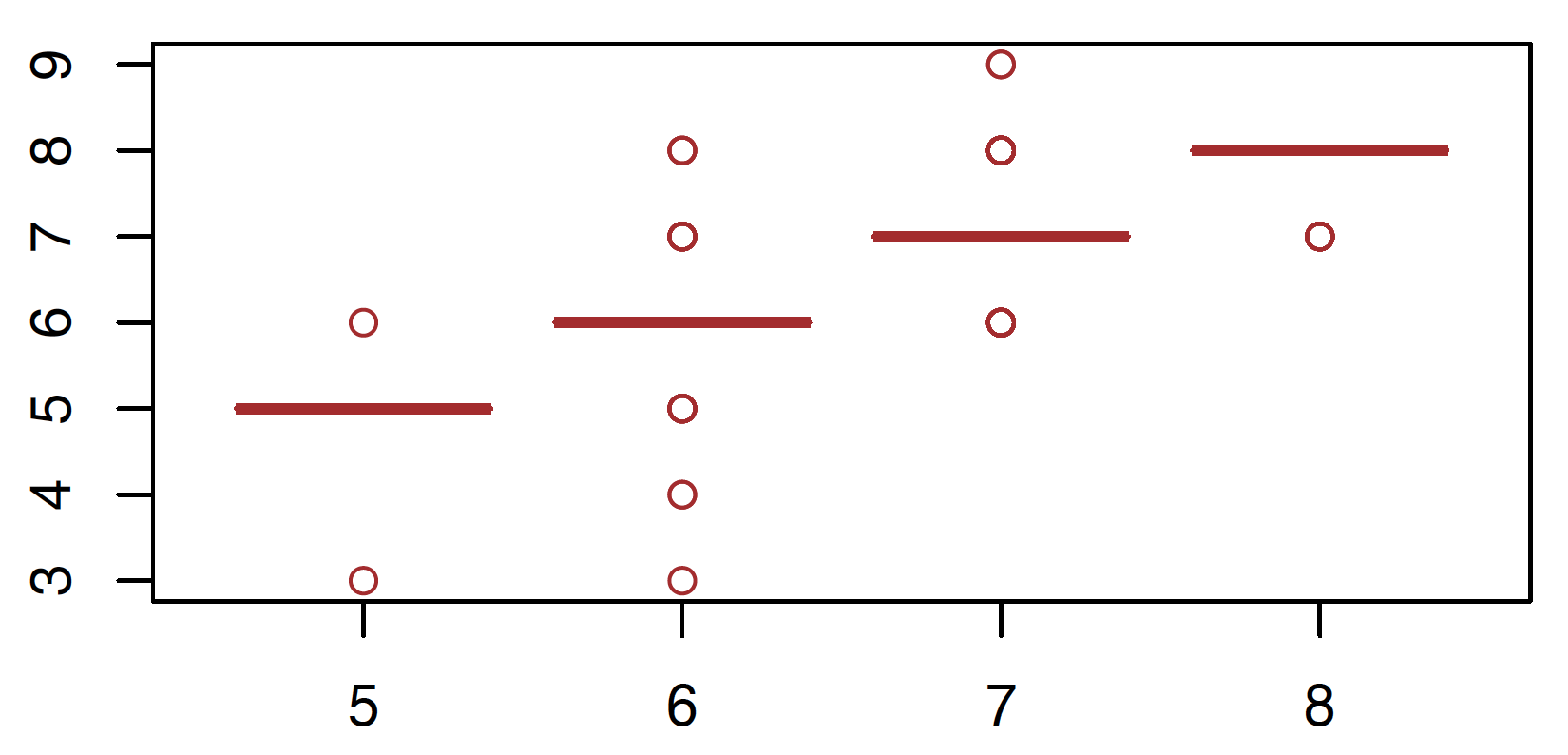}
\caption{The boxplot of the results for the training sample. The 568 samples are used to train the SVM predictive model. Thevalue of mood from 5 to 8 are predicted from 3 to 9.}
\label{fig:predict_1}
\end{figure}

\par
\begin{figure}[!ht]
\centering
\includegraphics[width=0.5\textwidth]{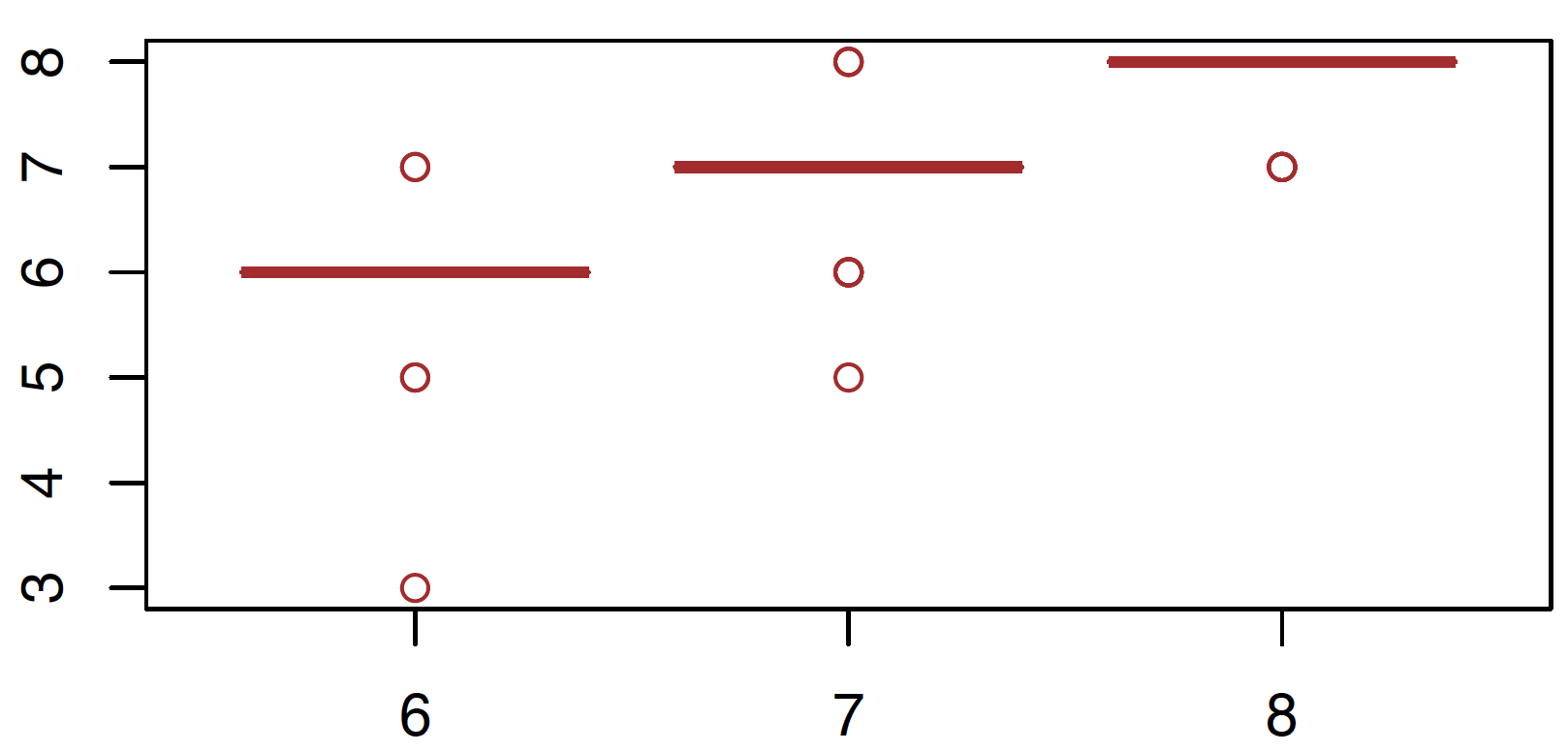}
\caption{The boxplot of the results that is tested for the test sample. The most predictive results are correct.}
\label{fig:predict_2}
\end{figure}

 Furthermore, we verified the accuracy of SVM predictive model by predicting the test sample. We have divide 100 sample as the testing sample in section \ref{sec:pre}. We have the result as shown in Table \ref{tab:test} and the statistical results in Fig. \ref{fig:predict_2}. The 81 samples were correctly predicted. 

\begin{table}[ht]
\centering
\begin{tabular}{p{2cm} p{1cm} p{1cm} p{1cm} p{1cm} p{1cm}}
\hline
result\_test & 3 & 5 & 6  & \textit{7}  & 8  \\ \hline
6            & 1 & 2 & 10 & \textit{2}  & 0  \\ 
7            & 0 & 1 & 5  & \textit{55} & 3  \\ 
8            & 0 & 0 & 0  & \textit{5}  & 16 \\ \hline
\end{tabular}
\caption{The predicted mood by SVM predictive model for the test sample. And the statistical results in Fig. \ref{fig:predict_2}.}
\label{tab:test}
\end{table}

Last, we test the accuracy of benchmark that assuming the mood of user is the same as the previous day, which is 62.3\%. Therefore, we have the comparison as shown in Table \ref{tab:com}.

\begin{table}[ht]
\centering
\caption{The comparison of testing based on training sample and testing sample, and the benchmark}
\label{tab:com}
\begin{tabular}{cccc}
\hline
prediction    & result\_train  & result\_test  &  benchmark \\ \hline
accuracy      & 0.822          & 0.810         & 0.623 \\ \hline
\end{tabular}
\end{table}

\subsection{Model Variant 2}

For this variant of the model, we incorporate a Recurrent Neural Network (RNN) to exploit the temporal characteristics of the dataset. To do so, we first pre-process the data somewhat. For this pre-processing we first replaced all the unavailable values, that is the values corresponding to `NA', by their values of the previous data-point. In such a way, we can use more data-point and do not have to discard any data points. Besides that, it seems reasonable to equate these values to their previously measured values since all the variables are measured several times a day, and it seems plausible that the values of these variables do not change substantially from one data-point to the next.

Now that we have a full dataset with no missing values, we can aggregate the data over the days. This allows us to obtain averages of all the days for each variable, which is needed to produce a prediction of the average mood the next day. At the same time, all the days that the mood variable is measured and delete the days in our dataset for which the mood is not observed for each individual separately. By doing this for each individual separately, we avoid throwing away data that we could in fact use for certain individuals. So if, for instance, the mood is only measured for an individual 1 at 10 dates, but for some other individual 2 on 15 dates, we avoid throwing away 5 dates for individual 2. As a next step, we then find for which days, where at least the mood has been measured, the most variables have been recorded and discard the rest of the days in our dataset. This is again done for each individual separately, which results in the following number of observations.

\begin{table}[ht!]
\centering
\caption{Number of observations after pre-processing the data}
\begin{tabular}{cccc}
\hline
\textbf{ID}      & \textbf{Observations} & \textbf{ID}      & \textbf{Observations} \\ \hline
\textit{AS14.01} & 18                    & \textit{AS14.19} & 21                    \\
\textit{AS14.02} & 11                    & \textit{AS14.20} & 12                    \\
\textit{AS14.03} & 17                    & \textit{AS14.23} & 11                    \\
\textit{AS14.05} & 21                    & \textit{AS14.24} & 18                    \\
\textit{AS14.06} & 12                    & \textit{AS14.25} & 14                    \\
\textit{AS14.07} & 14                    & \textit{AS14.26} & 15                    \\
\textit{AS14.08} & 19                    & \textit{AS14.27} & 14                    \\
\textit{AS14.09} & 9                     & \textit{AS14.28} & 13                    \\
\textit{AS14.12} & 15                    & \textit{AS14.29} & 14                    \\
\textit{AS14.13} & 17                    & \textit{AS14.30} & 14                    \\
\textit{AS14.14} & 11                    & \textit{AS14.31} & 12                    \\
\textit{AS14.15} & 12                    & \textit{AS14.32} & 13                    \\
\textit{AS14.16} & 14                    & \textit{AS14.33} & 11                    \\
\textit{AS14.17} & 15                    & \textit{}        &                       \\ \hline
\end{tabular}
\end{table}

As a final step for preparing to fit a RNN, we convert all variables to the $[0, 1]$ interval, which is necessary for the RNN to converge (faster). In the end, we scale back our predictions to their original scale such that we obtain predictions for the mood that we actually observe.

For every individual we then train and test a RNN, where we eventually ended with a learning rate of $7\%$, $4$ hidden layers in the network, $10.000$ iterations, the logistic sigmoid and the stochastic gradient descent method as updating rule. For testing the individual RNN's, we used $30\%$ of the available data (rounded to the nearest integer) and the other $70\%$ (rounded to the nearest integer) for training the data. As an example, we present the results of this training and testing phase for individual AS14.08 below. Note that for each individual, the random number generator in the training phase was initialized by \textsc{set.seed(2204)}.
\begin{figure}[!ht]
\caption{Training phase for AS14.08}
\begin{subfigure}[h]{0.6\textwidth}
    \centering
    \includegraphics[width=\textwidth]{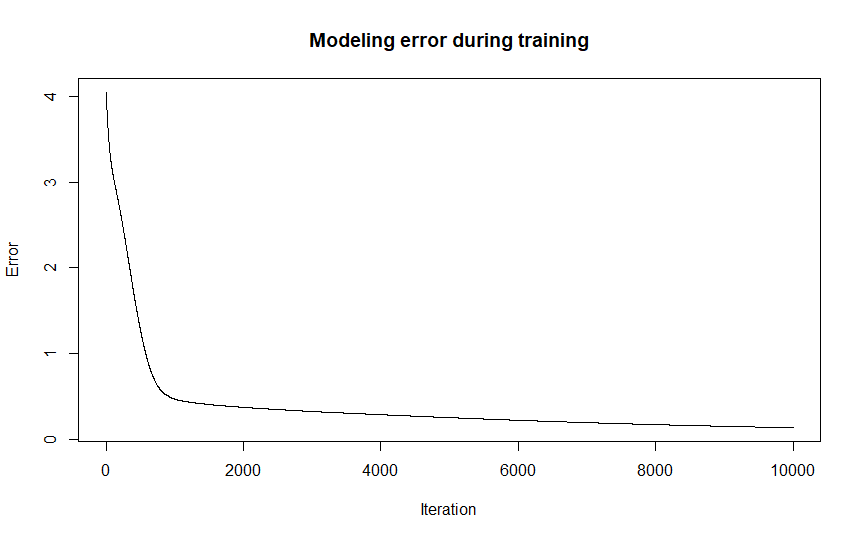}
	\caption{Error for training set of AS14.08}
\end{subfigure} 
\begin{subfigure}[h]{0.6\textwidth}
    \centering
    \includegraphics[width=\textwidth]{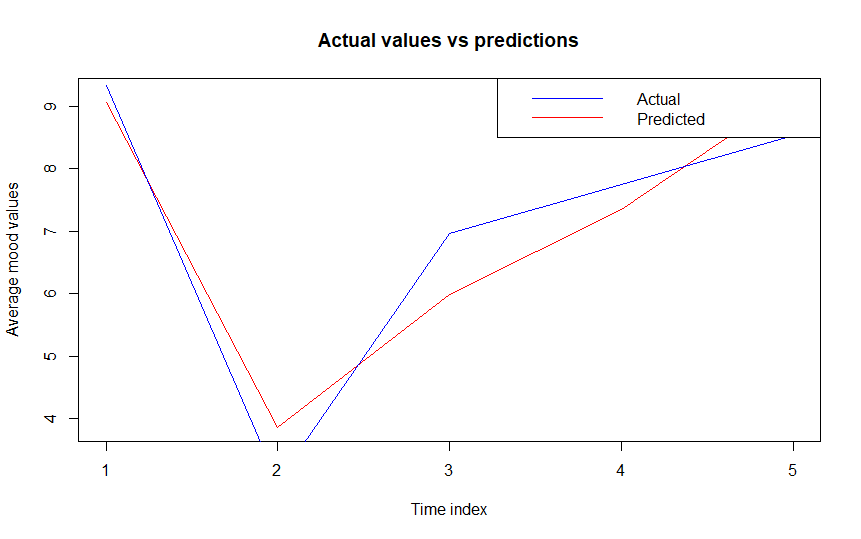}
	\caption{Predictions for test set of AS14.08}
    \end{subfigure}
\label{error_predictions}
\end{figure}

From results in \autoref{error_predictions}, the errors made in classifying the data decrease rather steep as the number of iterations progress. From the corresponding prediction plot, we see what the actual values of the mood were in the test set as opposed to the predicted values from the trained RNN. We might be worried from the error plot that we are overfitting the data in the RNN since the errors become so small, but we can see that the RNN seems to reasonably predict the mood for the following day from the prediction plot. This means our RNN is not overfitting in this case and that it can reasonably predict the mood for the following day for unknown cases. If we train our network using the entire dataset, we can also see that we adequately capture the mood of the following day for the known cases.

\begin{figure}[!htbp]
\centering
\includegraphics[scale=0.5]{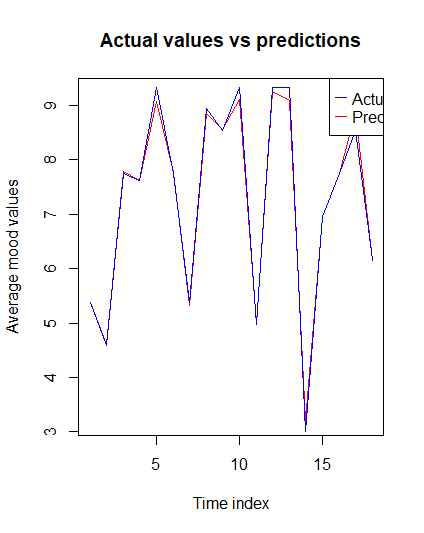}
\caption{Predictions for entire dataset of AS14.08}
\label{fig:predictions}
\end{figure}

The results of predictions in \autoref{fig:predictions} show that we actually capture the mood of the following day with rather high accuracy. As expected, we thus obtain qualitatively the same pattern for the errors made in classifying the data as for the training phase earlier arises when we use the entire dataset, explains why we are able to predict the mood of the following day quite precisely. Moreover, this pattern for both the predictions and the errors is consistent for all the individuals. To provide a selection of our results, we show the prediction plots for 3 individuals below, namely for AS14.08, AS14.16 and AS14.24.
\begin{figure}[htbp!]
\caption{Actual values vs predictions for each individual}
\begin{subfigure}[h]{0.34\textwidth}
    \centering
    \includegraphics[width=\textwidth]{Predictions_AS14_08.png}
    \caption{AS14.08}
    \end{subfigure} \hspace{-0.5cm}
\begin{subfigure}[h]{0.34\textwidth}
    \centering
    \includegraphics[width=\textwidth]{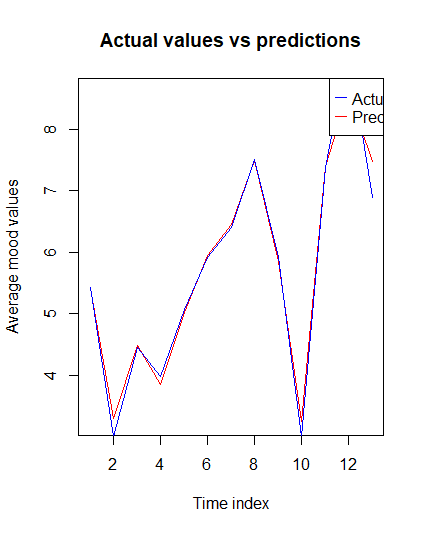}
    \caption{AS14.16}
    \end{subfigure} \hspace{-0.5cm}
\begin{subfigure}[h]{0.34\textwidth}
    \centering
    \includegraphics[width=\textwidth]{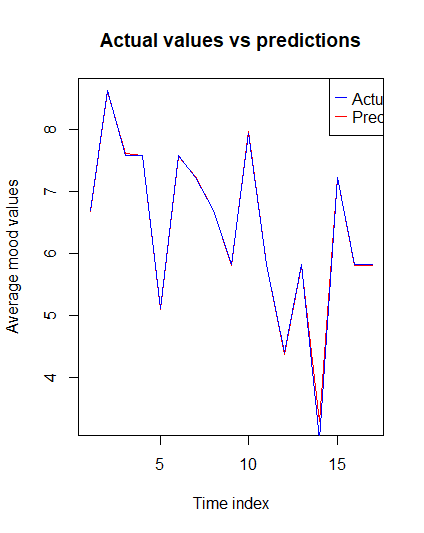}
    \caption{AS14.24}
    \end{subfigure}
\end{figure}
From these plots we indeed see that our predictions match the observed values rather closely and seem pretty accurate. This is confirmed when we inspect the RMSE of the predictions for each individual. These RMSE's are given in the following table.
\begin{table}[ht!]
\centering
\setlength\tabcolsep{6pt} % default value: 6pt
\caption{RMSE of RNN approach}
\begin{tabular}{cccc}
\hline
\textbf{ID}      & \textbf{RMSE} & \textbf{ID}      & \textbf{RMSE} \\ \hline
\textit{AS14.01} & 0.4013390                    & \textit{AS14.19} & 0.4711004                    \\
\textit{AS14.02} & 0.2142030                    & \textit{AS14.20} & 0.3762178                    \\
\textit{AS14.03} & 0.4265163                    & \textit{AS14.23} & 0.2910442                    \\
\textit{AS14.05} & 1.0045396                    & \textit{AS14.24} & 0.2965332                    \\
\textit{AS14.06} & 0.2778267                    & \textit{AS14.25} & 0.3889919                    \\
\textit{AS14.07} & 0.2246084                    & \textit{AS14.26} & 0.2770717                    \\
\textit{AS14.08} & 0.5791765                    & \textit{AS14.27} & 0.3803674                    \\
\textit{AS14.09} & 0.5235971                     & \textit{AS14.28} & 0.2794391                    \\
\textit{AS14.12} & 0.3836072                    & \textit{AS14.29} & 0.5301942                    \\
\textit{AS14.13} & 0.4786648                    & \textit{AS14.30} & 0.2453128                    \\
\textit{AS14.14} & 0.2647090                    & \textit{AS14.31} & 0.3185494                    \\
\textit{AS14.15} & 0.2972429                    & \textit{AS14.32} & 0.2421542                    \\
\textit{AS14.16} & 1.0184591                    & \textit{AS14.33} & 0.3605297                    \\
\textit{AS14.17} & 0.4683970                    & \textit{}        &                       \\ \hline
\end{tabular}
\end{table}
We see that these RMSE's are all pretty close to zero for each individual. All things considered, we thus see that the predicted values match the actual values rather closely, for each individual. The RNN method thus seems to adequately incorporate the temporal aspects of the dataset at hand on an individual level.

\subsection{Model Variant 3}

In this model variant, we simply predict that the average mood on the next day is the same as on this day. In the prediction plots below we can see these actual and predicted values for 3 individuals, namely for AS14.08, AS14.16 and AS14.24.
\begin{figure}[!htbp]
\setlength\tabcolsep{6pt} % default value: 6pt
\caption{Actual values vs predictions for each individual (1)}
\begin{subfigure}[h]{0.7\textwidth}
    \centering
    \includegraphics[width=\textwidth]{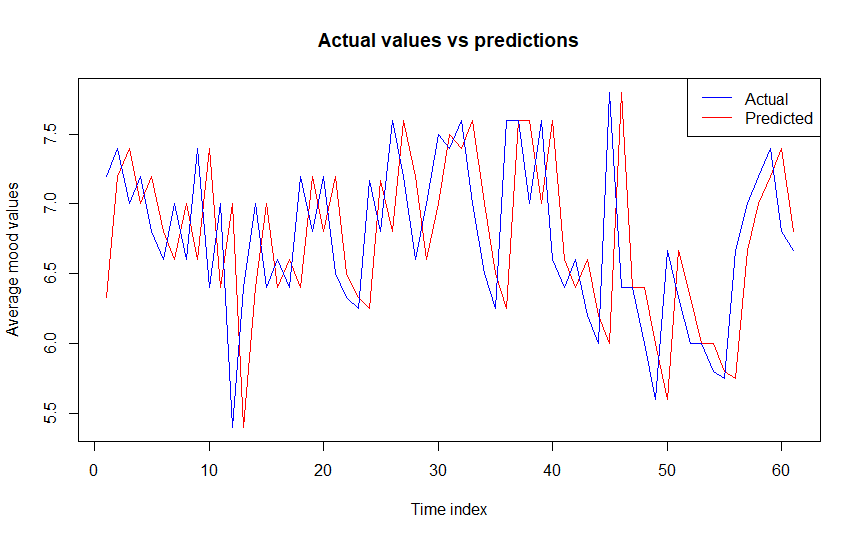}
    \caption{AS14.01}
\end{subfigure} \hspace{-0.5cm}
\begin{subfigure}[h]{0.7\textwidth}
    \centering
    \includegraphics[width=\textwidth]{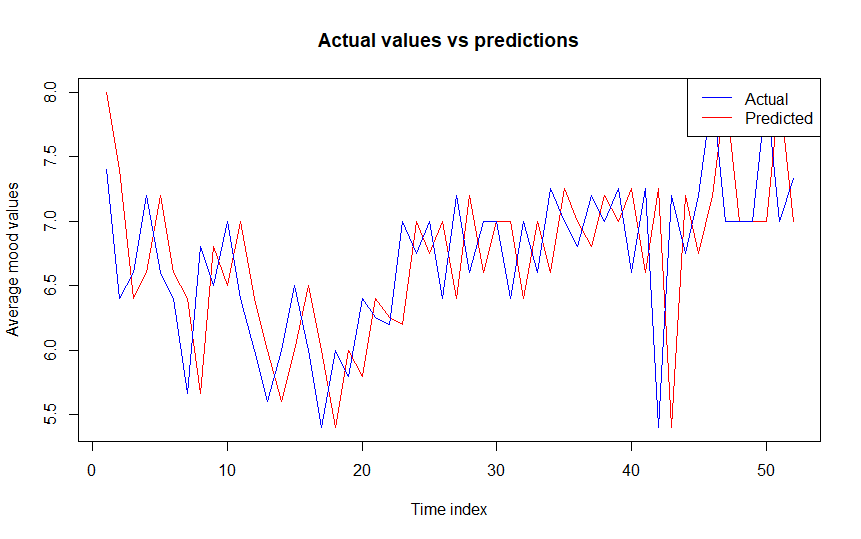}
    \caption{AS14.02}
    \end{subfigure}
\begin{subfigure}[h]{0.7\textwidth}
    \centering
    \includegraphics[width=\textwidth]{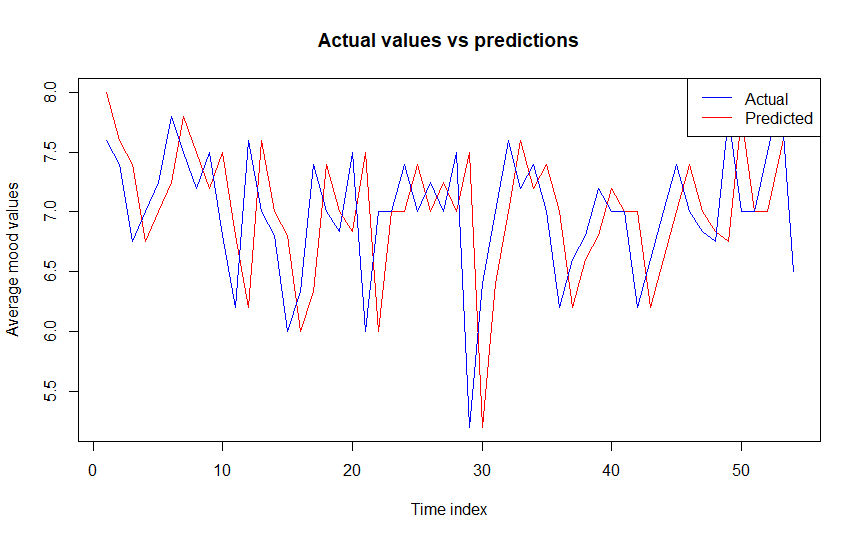}
    \caption{AS14.03}
    \end{subfigure}
\end{figure}
We can see that this naive approach of simply predicting that the average mood will stay constant (the same as the previous day) does not produce as nice results as those of the RNN's. The following table shows the corresponding RMSE for each individual when adopting this naive approach.

% \begin{table}[ht!]
% \centering
% \setlength\tabcolsep{3pt} % default value: 6pt
% \caption{RMSE of naive approach}
% \begin{tabular}{cccccccccccc}
% \hline
% \textbf{ID}     & \textit{AS14.01} &\textit{AS14.02}&\textit{AS14.03}&\textit{AS14.05}&\textit{AS14.06}&\textit{AS14.07}&\textit{AS14.08}&\textit{AS14.09}&\textit{AS14.12}&\textit{AS14.13}&\textit{AS14.14}&\textit{AS14.15}&\textit{AS14.16}&\textit{AS14.17} \\
% \textbf{RMSE}    & 3.802594  & 6.134013  & 2.906315  & 5.314184  & 4.985479  & 10.497090  & 4.977672   & 5.282255 & 4.506662  & 7.353911  & 4.541047  & 3.139621  & 4.735328  & 3.596140 \\
% \textbf{ID}     & \textit{AS14.19} & \textit{AS14.20} & \textit{AS14.23}& \textit{AS14.24}  & \textit{AS14.25} & \textit{AS14.26} & \textit{AS14.27} & \textit{AS14.28} & \textit{AS14.29} & \textit{AS14.30} & \textit{AS14.31} & \textit{AS14.32} & \textit{AS14.33} &  \\
% \textbf{RMSE}    & 4.642377 & 3.289039 & 3.544714 & 5.064912 & 3.896437 & 6.547519 & 5.096676 & 5.410381 & 4.377468 & 2.966854 & 3.009430 & 4.637708 & 7.052462 &   \\ \hline
% \end{tabular}
% \end{table}

\begin{table}[ht!]
\centering
\setlength\tabcolsep{6pt} % default value: 6pt
\caption{RMSE of naive approach}
\begin{tabular}{cccc}
\hline
\textbf{ID}      & \textbf{RMSE} & \textbf{ID}      & \textbf{RMSE} \\ \hline
\textit{AS14.01} & 3.802594                    & \textit{AS14.19} & 4.642377                    \\
\textit{AS14.02} & 6.134013                    & \textit{AS14.20} & 3.289039                    \\
\textit{AS14.03} & 2.906315                    & \textit{AS14.23} & 3.544714                    \\
\textit{AS14.05} & 5.314184                    & \textit{AS14.24} & 5.064912                    \\
\textit{AS14.06} & 4.985479                    & \textit{AS14.25} & 3.896437                    \\
\textit{AS14.07} & 10.497090                    & \textit{AS14.26} & 6.547519                    \\
\textit{AS14.08} & 4.977672                    & \textit{AS14.27} & 5.096676                    \\
\textit{AS14.09} & 5.282255                     & \textit{AS14.28} & 5.410381                    \\
\textit{AS14.12} & 4.506662                    & \textit{AS14.29} & 4.377468                    \\
\textit{AS14.13} & 7.353911                    & \textit{AS14.30} & 2.966854                    \\
\textit{AS14.14} & 4.541047                    & \textit{AS14.31} & 3.009430                    \\
\textit{AS14.15} & 3.139621                    & \textit{AS14.32} & 4.637708                    \\
\textit{AS14.16} & 4.735328                    & \textit{AS14.33} & 7.052462                    \\
\textit{AS14.17} & 3.596140                    & \textit{}        &                       \\ \hline
\end{tabular}
\end{table}
From this table we also see that the RNN's actually perform much better than the naive approach. All things considered, the naive approach does not seem correct to adopt and can be considered as a `clueless' method, that is if we had no idea how to approach the problem then this would be the standard `worst case scenario' for producing predictions. The naive approach can therefore indeed be considered as a benchmark model.

%%%%%%%%%%%%%%%%%%%%%%%%%%%%%%%%%%%%%%%%%%%
\section{Conclusion}
\label{sec:conclusion}
%%%%%%%%%%%%%%%%%%%%%%%%%%%%%%%%%%%%%%%%%%%
Hotel recommendation system is a popular research field. This paper provide a comprehensive pipeline for the researchers to build such a system from the raw data to specific application. Although the results show that the two methods achieve a successful prediction system, they are only the basic approaches in machine learning. Many approaches are interesting to further exploration. For instance, evolutionary approaches have been applied in many areas \cite{lan2020time}. Neuroevolution have been applied in evolving neural network for real-time computer vision \cite{lan2019evolving}, evolutionary robotics \cite{lan2019simulated,lan2019learning,lan2019evolutionary,lan2018directed}. In many areas \cite{lan2016convolution}, convolution neutral networks generally achieves remarkable performance that we aim to apply in this pipeline. Knowledge graph is a popular method that is applied to the many applications \cite{Liu2020Influence,liu2019evidence}, such as finance, medicine, biology, Question—Answering, Storing Information of Research, in particular recommendation system. Therefore, we will use knowledge graph to design the hotel recommendation system in the future. In addition, the signal compress \cite{lan2016bayesian,lan2017development,lan2016development} is an interesting technology for the pre-process raw data. These approaches are the points we aim to extend for this pipeline. 

\bibliographystyle{unsrt}
\bibliography{datamining}

\end{document}